\documentclass{elsart}

\usepackage{natbib}
\usepackage{graphicx}

\usepackage{amssymb}

\begin{document}

\begin{frontmatter}



\title{The Twin-Jet of NGC\,1052 at Radio, Optical, and X-Ray Frequencies}


\author[label1]{M.\,Kadler,}
\author[label1]{E.\,Ros,}
\author[label2]{J.\,Kerp,}
\author[label1]{H.\,Falcke,}
\author[label1]{J.\,A.\,Zensus,}
\author[label3]{R.\,W.\,Pogge,}
\author[label4]{G.\,V.\,Bicknell}

\address[label1]{Max-Planck-Institut f\"ur Radioastronomie, Auf dem H\"ugel 69, D-53121 Bonn, Germany}

\address[label2]{Radioastronomisches Institut der Universit\"at Bonn, Auf dem H\"ugel 71, D-53121 Bonn, Germany}

\address[label3]{Department of Astronomy, Ohio State University, 140 West 18th Avenue, Columbus, OH 43210-1173, U.S.A.}

\address[label4]{Research School of Astronomy \& Astrophysics, Mt. Stromlo Observatory, Cotter Road, Weston, ACT 2611, Australia}

\begin{abstract}
We present results from a combined radio, optical, and X-ray study of the
jet-associated emission features in NGC\,1052. We analyse the
radio-optical morphology and find a good
positional correlation between the radio jet and the optical emission cone. 
Two optical emission knots are directly 
associated with radio counterparts
exhibiting a radio to X-ray broadband
spectrum not compatible with synchrotron emission. 
We discuss the possibility that the thermal soft spectrum of the extended X-ray emission
originates from 
jet driven shocks produced in the interaction between the jet-plasma and its
surrounding medium. 
\end{abstract}




\end{frontmatter}

\section{Introduction}
\label{intro}
The nearby\footnote{At the distance of NGC\,1052 of about $22.6$\,Mpc, 
1\,arcsec corresponds to $\sim$110\,pc. 
Throughout the paper, we use $H_0 = 65$\,km\,s$^{-1}$\,Mpc$^{-1}$.} 
low-luminosity AGN at the center of NGC\,1052 bears a symmetric 
twin-jet whose parsec-scale properties have been studied in great detail using
Very Long Baseline Interferometry (VLBI) observations (e.g., Vermeulen et al. \citeyear{Ver02},
Kadler et al. \citeyear{Kad02a}). These studies have revealed the presence of a
dense circumnuclear absorber covering the inner parsecs of both jets.
On larger scales the source shows a radio structure with a very dominant 
compact core and two lobes of extended emission spanning about 15\,arcsec 
east and west of the core (Wrobel \citeyear{Wro84}). 
Based on {\it CHANDRA} data \citet{Kad02b} reported the detection of an associated X-ray jet 
with the spectral signature of a 0.5\,keV thermal plasma.  
While the diffuse, extended soft X-ray emission tends to anti-correlate with
the kiloparsec-scale radio emission of the lobes, two emission knots east
and west of the nucleus coincide positionally with features of
the radio jet. In the optical regime, \citet{Pog00} detected a conical structure 
associated with the nucleus as well as emission knots east and
west of the core in a {\it Hubble Space Telescope} H$\alpha$ image
of NGC\,1052. 

\section{Radio-optical morphology}
\label{radio-optical}
\begin{figure}[p!]
\centering
\includegraphics[width=12cm,clip]{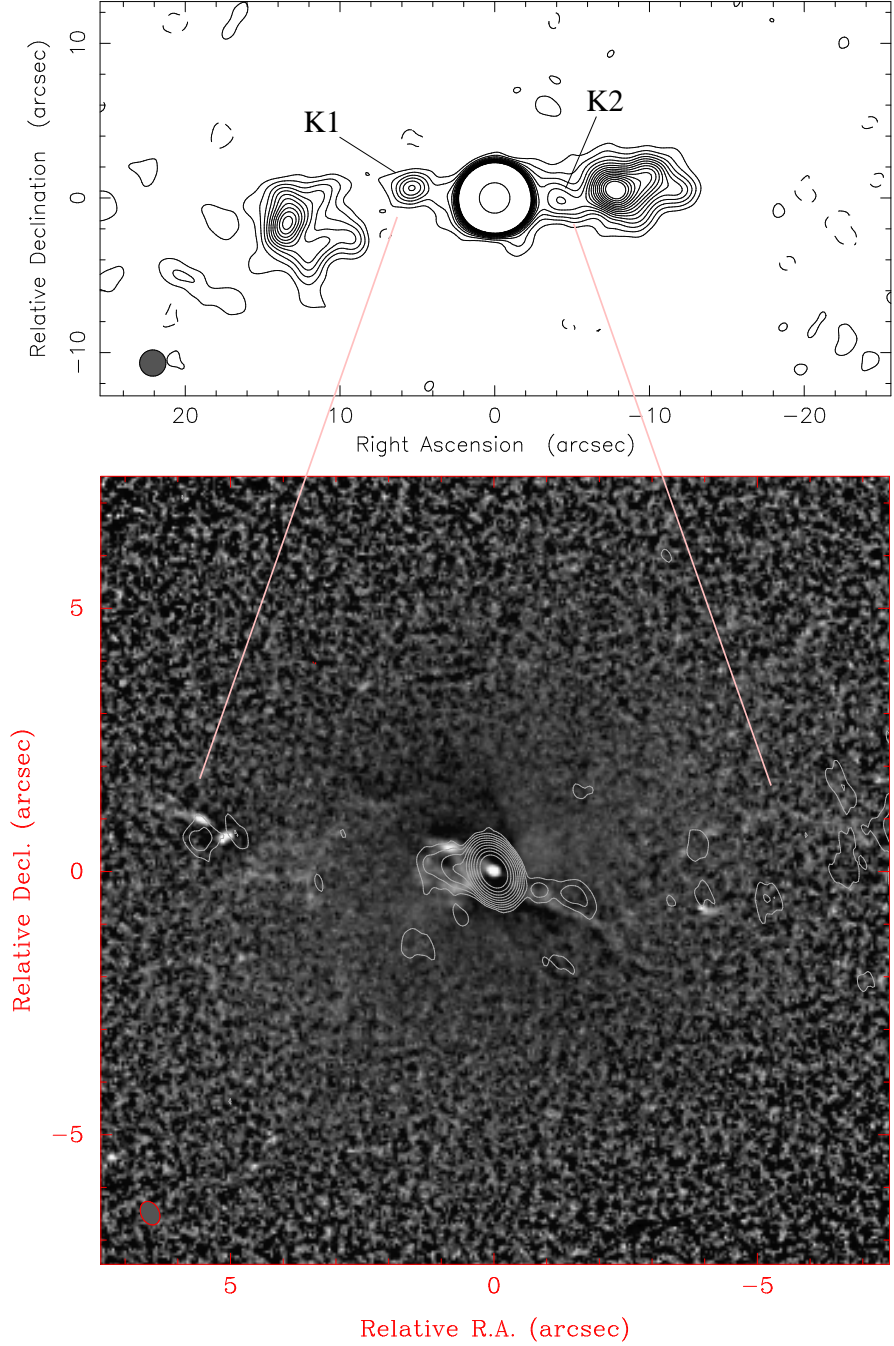}
\caption{
Top: Tapered (circular beam of 1.7\,arcsec FWHM), naturally weighted MERLIN map of NGC\,1052 at 1.4\,GHz with contours shown for
(-1, 1, 2, 3, 4, 5, 6, 7, 8, 9, 10, 11, 12, 13, 14, 15, 800) $\times$ 0.5\,mJy/beam. 
Bottom: Structure map for NGC\,1052 in the F\,658\,N filter. Dark regions represent dust obscuration
while bright regions are locations of enhanced emission. Overlaid is
the pure naturally weighted MERLIN map showing the
core of NGC\,1052 with
sub-arcsecond resolution (beam FWHM: 0.47$\times$0.34\,arcsec at 27.5$^\circ$). 
Contours at ($-1$, 1, 2, 4, 8, 16, 32, 64, 128, 256, 512, 1024) $\times$ 0.5\,mJy/beam are
shown. 
} 
\label{fig:merge}
\end{figure}

``Structure mapping'' \citet{Pog02} is an improved method of image contrast 
enhancement allowing to study the distribution of absorbing and emitting
regions in the circumnuclear regions of galaxies. 
For comparison of the optical and radio structure of NGC\,1052 we
analyzed archival MERLIN\footnote{MERLIN is a National Facility operated by the University of Manchester at Jodrell Bank Observatory on behalf of PPARC. The
observation was planned and scheduled by Dr.\,A.\,Pedlar.} data
from an observation at 1.4\,GHz on November 22, 1995.
The data post-processing was done applying standard methods within {\sc difmap}.
The resulting tapered MERLIN image is shown
in Fig.~\ref{fig:merge} together with
a structure map of NGC\,1052
based on the same observational data as used by \citet{Pog00}.
Superimposed are contours of the untapered full resolution MERLIN image.
Jet and counterjet extend continuously out to $\sim$1.5\,arcsec with
a position angle similar to the pc-scale twin-jet. 
The jet-counterjet structure is aligned with the optical emission
cone. The dark band perpendicular to the radio jet might be an artifact of the
image processing or might alternatively represent the signature of an
obscuring dusty region. 
The two optical emission knots in the east are located at the
edges of two radio sub-components of knot K1. The optical emission knot in the
west coincides roughly with a weak ($\sim 1 \sigma$) radio feature while the
stronger radio knot $1.5$\,arcsecond further out (K\,2) has no corresponding bright 
optical counterpart. Both knots, however, coincide positionally with an
uncertainty of $\sim$4\,arcsec with X-ray knots in the jet structure (see
Kadler et al. \citeyear{Kad02b}).
The origin of the optical emission remains unclear since there is no
continuum image to subtract from the H$\alpha$
filter image of NGC\,1052. The morphological similarity of the emission cone
in this LINER 1.9 galaxy
to the features typically observed in Seyfert\,2 galaxies (Falcke et al. \citeyear{Fal98}),
however, suggests an origin in the narrow emission line region. 
The optical flux density of the two eastern knots of ($68\pm1$)\,$\mu$Jy 
exceeds the power-law extrapolation from the radio to the X-ray regime by
almost three orders of magnitude ruling out synchrotron emission as a
possible mechanism.

\section{X-ray emission from jet-driven shocks}

The above data provide circumstantial evidence that shocks are
implicated in the production
of the jet-associated X-ray emission, 
produced as a result of the dissipation of some
fraction of the jets kinetic power in jet-cloud collisions.
Given a 1.4\,GHz
flux density of 0.15\,Jy of both radio lobes, 
the corresponding radio luminosity is $L_{1.4}
\sim 10^{29}$\,erg\,s$^{-1}$\,Hz$^{-1}$. 
Using the expressions given in \citet{bicknell98a}
the  kinetic power in
both jets is $2F_E \sim
\kappa_{1.4}^{-1} L_{1.4}$, where
$\kappa_\nu \sim 10^{-14}-10^{-10}$\,Hz$^{-1}$ is the ratio of monochromatic
radio power to kinetic power.
Thus the kinetic power per jet is $F_E \sim 5 \cdot 10^{41} (\kappa_{1.4} / 10^{-13})^{-1}$\,erg\,s$^{-1}$. 
The 0.3--8\,keV X-ray luminosity associated
with the jets is $L_{X,\rm obs}
\sim 1.2 \cdot 10^{40}$\,erg\,s$^{-1}$, i.e., approximately 2.5\%
of the estimated jet energy flux. This is
expected since shocks driven into dense clouds by a light jet only
tap a small fraction of
the jet power. 
The
energy flux into the shock is given by
$F_E \cdot f(M_{\rm jet}, \Delta \theta) \cdot (v_{\rm sh}/v_{\rm jet})$, 
where $v_{\rm sh}$ is the shock velocity and
the numerical factor $f(M_{\rm jet}, \Delta \theta)$ depends
upon the jet Mach number,
$M_{\rm jet}$, and the deflection angle, $\Delta \theta$, but is
mainly sensitive to the latter. It
is approximately 0.2 for a $10^\circ$ deflection and 1 for a
head-on collision. 
The immediate post-shock temperature is $T_{\rm sh}
\sim (3 \mu m_p/16k)
v_{\rm sh}^2$ where $m_p$ is the proton mass and $\mu$ is the mean
molecular weight of the shocked particles, so that 0.5\,keV corresponds to a shock velocity,
$v_{\rm sh} \sim 640$\,km\,s$^{-1}$. 
In a radiative shock,
approximately half of the energy flux into
the shock is radiated in the hot continuum so that the
expected X-ray luminosity is $L_X
\sim 1.7 \cdot 10^{39} f (v_{\rm sh} /10^3 \, {\rm km \>s^{-1}})
(v_{\rm jet}/c)^{-1} \, (2 F_E/ 10^{42} \> \rm erg \> s^{-1})$\,erg\,s$^{-1}$.
This estimate is compatible to the observed value
for a
mildly relativistic jet with $v_{\rm jet} \sim 0.25 c$ (see Vermeulen
et al. \citeyear{Ver02})
if $f \sim 1$, which would imply 
the complete disruption of the 
jet rather than a glancing deflection.





\begin{thebibliography}{}


\bibitem[Bicknell et al. (1998)]{bicknell98a} Bicknell, G.~V., Dopita, M.~A.,
Tsvetanov, Z.~I., \& Sutherland, R.~S.
1998, ApJ, {495}, 680

\bibitem[Falcke et al. (1998)]{Fal98} Falcke, H., Wilson, A. S., Simpson, C. 1998, ApJ, 502, 199

\bibitem[Kadler et al. (2002)]{Kad02b} Kadler, M., Ros, E., Kerp, J., Lobanov, A.
P., Falcke, H., Zensus, J. A., Proceedings of the 6th European VLBI Network Symposium,
     Ros, E., Porcas, R.W., Lobanov, A.P., Zensus, J.A. (eds.) 2002, Bonn, Germany, p. 167

\bibitem[Kadler et al. (2003)]{Kad02a} Kadler, M., Ros, E., Zensus, J. A., Lobanov, A. P., Falcke, H. 2003, SRT Conference Proceedings, Vol. 1, in press (astro-ph/0204054)

\bibitem[Pogge et al. (2000)]{Pog00} Pogge, R. W., Maoz, D., Ho, L. C., Eracleous, M. 2000, ApJ, 532, 323

\bibitem[(Pogge et al. 2002)]{Pog02} Pogge, R. W., Martini, P. 2002, ApJ, 569, 624

\bibitem[Vermeulen et al. (2003)]{Ver02} Vermeulen, R. C., Ros, E., Kellermann, K. I., Cohen, M. H., Zensus, J. A., van Langevelde, H. J., 2003, A\&A, in press 

\bibitem[Wrobel (1984)]{Wro84} Wrobel, J. M., ApJ, 1984, 284, 531

\end{thebibliography}
\end{document}